\documentclass{ptephy_v1}

\preprintnumber{XXXX-XXXX} 
\usepackage[driverfallback=dvipdfm]{hyperref}
\usepackage{orcidlink}
\usepackage{authblk-TI}

\usepackage{natbib}

\begin{document}

\title{Input optics systems of the KAGRA detector during O3GK}












\author{
T.~Akutsu\,\orcidlink{0000-0003-0733-7530}$^{1,2}$, 
M.~Ando$^{3,4,1}$, 
K.~Arai$^{5}$, 
Y.~Arai$^{5}$, 
S.~Araki$^{6}$, 
A.~Araya\,\orcidlink{0000-0002-6884-2875}$^{7}$, 
N.~Aritomi\,\orcidlink{0000-0003-4424-7657}$^{3}$, 
H.~Asada\,\orcidlink{0000-0001-9442-6050}$^{8}$, 
Y.~Aso\,\orcidlink{0000-0002-1902-6695}$^{9,10}$, 
S.~Bae\,\orcidlink{0000-0003-2429-3357}$^{11}$
% ...
\\
(The KAGRA Collaboration), \\
R.~Goetz$^{78}$,       
M.~Heintze$^{82}$,    
J.~Liu$^{83}$,             
C.~M\"{u}ller$^{78}$,    
R.~L.~Savage$^{84}$,     
and 
D.~Tanner$^{78}$       
}

\affil{
$^{1}$Gravitational Wave Science Project, National Astronomical Observatory of Japan, 2-21-1 Osawa, Mitaka City, Tokyo 181-8588, Japan\\
$^{2}$Advanced Technology Center, National Astronomical Observatory of Japan, 2-21-1 Osawa, Mitaka City, Tokyo 181-8588, Japan\\
$^{3}$Department of Physics, The University of Tokyo, 7-3-1 Hongo, Bunkyo-ku, Tokyo 113-0033, Japan\\
$^{4}$Research Center for the Early Universe, The University of Tokyo, 7-3-1 Hongo, Bunkyo-ku, Tokyo 113-0033, Japan\\
$^{5}$Institute for Cosmic Ray Research, KAGRA Observatory, The University of Tokyo, 5-1-5 Kashiwa-no-Ha, Kashiwa City, Chiba 277-8582, Japan\\
$^{6}$Accelerator Laboratory, High Energy Accelerator Research Organization (KEK), 1-1 Oho, Tsukuba City, Ibaraki 305-0801, Japan\\
$^{7}$Earthquake Research Institute, The University of Tokyo, 1-1-1 Yayoi, Bunkyo-ku, Tokyo 113-0032, Japan\\
$^{8}$Department of Mathematics and Physics, 
Graduate School of Science and Technology, Hirosaki University, 3 Bunkyo-cho, Hirosaki, Aomori 036-8561, Japan\\
$^{9}$Kamioka Branch, National Astronomical Observatory of Japan, 238 Higashi-Mozumi, Kamioka-cho, Hida City, Gifu 506-1205, Japan\\
$^{10}$The Graduate University for Advanced Studies (SOKENDAI), 2-21-1 Osawa, Mitaka City, Tokyo 181-8588, Japan\\
$^{11}$Korea Institute of Science and Technology Information, 245 Daehak-ro, Yuseong-gu, Daejeon 34141, Republic of Korea\\
% ...
$^{78}$Department of Physics, University of Florida, Gainesville, FL 32611, USA\\
% ...
$^{82}$LIGO Livingston Observatory, Livingston, LA 70754, USA\\
$^{83}$OzGrav, University of Western Australia, Crawley, Western Australia 6009, Australia\\
$^{84}$LIGO Hanford Observatory, Richland, WA 99352, USA
}




\begin{abstract}%
KAGRA, the underground and cryogenic gravitational-wave detector, was operated for its solo observation from February 25th to March 10th, 2020, and its first joint observation with the GEO 600 detector from April 7th -- 21st, 2020 (O3GK). This study presents an overview of the input optics systems of the KAGRA detector, which consist of various optical systems, such as a laser source, its intensity and frequency stabilization systems, modulators, a Faraday isolator, mode-matching telescopes, and a high-power beam dump. These optics were successfully delivered to the KAGRA interferometer and operated stably during the observations.
The laser frequency noise was observed to limit the detector sensitivity above a few kHz, whereas the laser intensity did not significantly limit the detector sensitivity.
\end{abstract}

\subjectindex{F30}

\maketitle

\section{Introduction}
Approximately 100 years after Einstein predicted (in 2015) gravitational waves (GWs)
as a consequence of general relativity, two advanced laser interferometer gravitational-wave observatory (aLIGO) detectors observed a binary black hole merger in their first observing run, O1 \cite{Abbott:2016blz}. In their second observing run, a binary neutron star merger was detected \cite{Abbott:2018exr}. Fortunately, at that time, a third detector, Advanced Virgo (AdV), was online in addition to the two aLIGO detectors. This enabled better sky localization; therefore, other types of observatories were successful in identifying the corresponding object \cite{GBM:2017lvd}. This observation, together with GW detectors and other various telescopes, yielded various astrophysical and cosmological insights \cite{Abbott:2017xzu}, successfully establishing multi-messenger astrophysics.

KAGRA \cite{Akutsu:2017kpk} is the GW detector built in Japan,
which features an underground site and cryogenic technologies.
Adding a fourth km-scale detector to the detector network will improve
the sky localization \cite{Wen:2010cr}, parameter estimation \cite{Aasi:2013wya},
and detectability of GW polarization \cite{Takeda:2018uai}.
Ten years after the KAGRA project commenced, the first observation run was conducted
from February 25th to March 10th, 2020 with a typical observable distance of 400 kpc for binary neutron star inspirals (BNS inspiral range \cite{Finn:1993}).
After a 3-week commissioning break, 
the observation was restarted with a typical sensitivity of 600 kpc
from April 7th -- 21st, 2020.
The latter was a joint observation with GEO 600 \cite{Abadie:2011xta}
in Germany, called the O3GK observation.
Our initial plan was to join the LIGO-VIRGO observation network
after KAGRA achieved a 1 Mpc binary neutron star range.
However, because the LIGO-VIRGO observation
was suspended on March 23rd, 2020, owing to the COVID-19 pandemic,
a joint observation could not be conducted with the LIGO-VIRGO detectors.

\subsection{Input optics systems}
\begin{figure}[t]
\begin{center}
\includegraphics[scale=0.5]{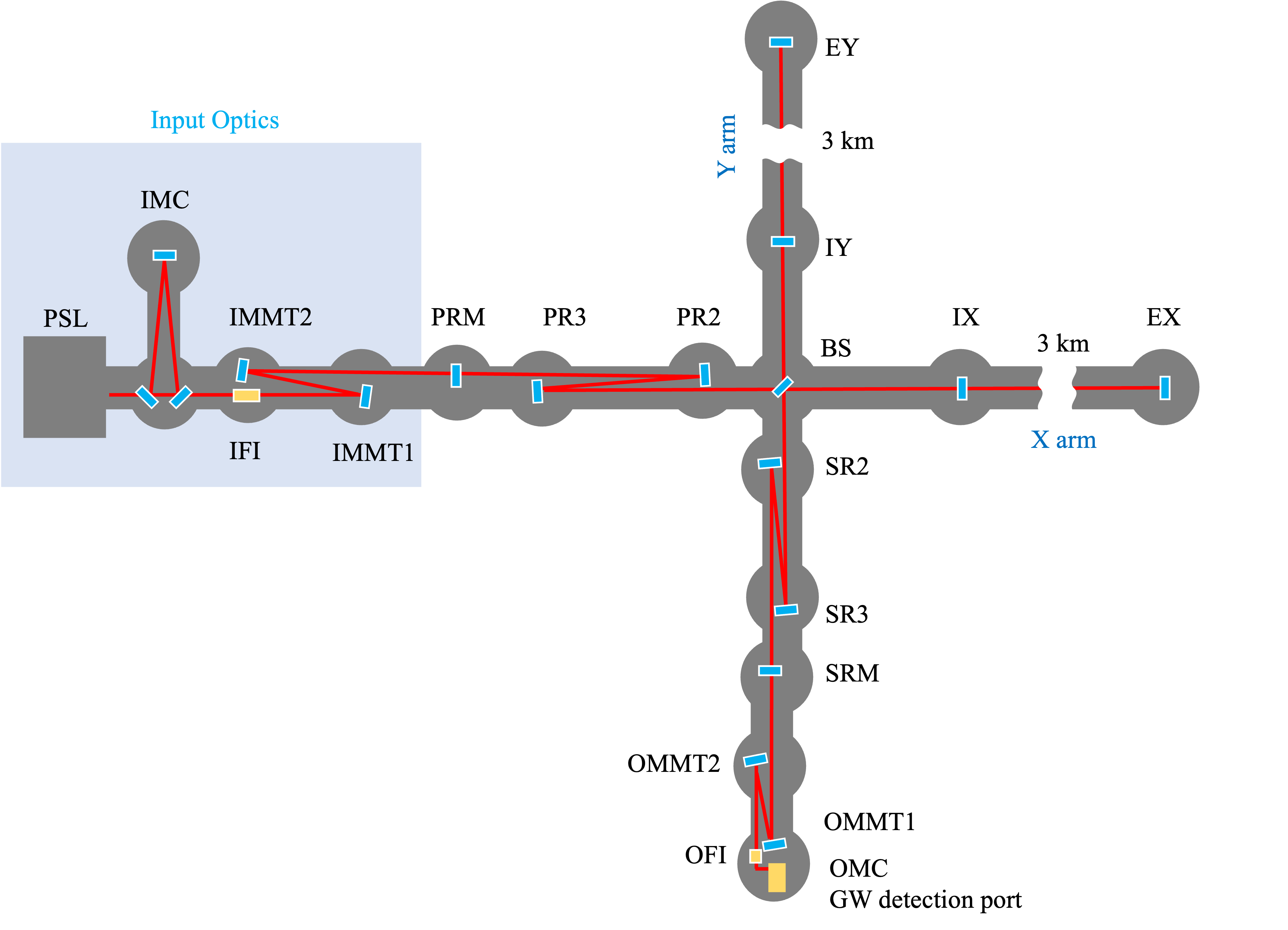}
\caption{\label{fig:ifoconfig}Optical configuration of KAGRA. The X and Y arms are each 3 km in length, forming Fabry-Perot cavities. The GW signals are detected at the transmission port of the output mode cleaner (OMC). The input optics system includes the PSL, IMC, IFI, and IMMTs. All the main optical components are in vacuum enclosures placed in the underground tunnel.
PRM, PR2, PR3: first, second, and third power-recycling mirrors;
BS: beam splitter; IX, EX, IY, EY: input or end X mirrors, input or end Y mirrors; SRM, SR2, SR3: first, second, and third signal-recycling mirrors;
OMMT: output mode-matching telescope. OFI: output Faraday isolator. The SRM was misaligned not to form the signal-recycling cavity during O3GK.}
\end{center}
\end{figure}

GW detectors are based on L-shaped Michelson laser interferometers.
The effects of GWs distorting space-time are detected
as differential length changes between the two arms.
Because a length change owing to GWs is extremely small,
typically in the order of 10$^{-21}$ in strain, 
attaining such sensitivity is extremely challenging.
The sensitivity of a detector is limited by various types of technical and fundamental noises.
To attain such a high sensitivity requires months, or even a year, of commissioning efforts.
That is why the first observation of KAGRA was limited to the BNS range of 400 kpc,
whereas the design sensitivity is 140 Mpc.
Detailed descriptions of the detector and its noise analysis during O3GK
are presented in Ref. \cite{KAGRA:2022fgc}.

The laser source and related input optics are among the most important
subsystems of the interferometer
because the properties of the laser directly affect the sensitivity
and stability of the main interferometer.
The goal of the input optics system of KAGRA is to provide
minimal noise and stable light to the main interferometer.
This study describes various components of the input optics systems
developed for the first observing run of KAGRA.
In the next section, an overview of the KAGRA interferometer
and the input optics systems will be presented.
Sec.~\ref{sec:PSL} to Sec.~\ref{sec:IMMT}
will review each input optics component.
In Sec.~\ref{sec:noise},
the contribution of laser intensity and frequency noise to the GW sensitivity will be discussed.
Finally, in Sec.~\ref{sec:plans}, prospects for the next observing run will be summarized.

\section{Input optics overview}

\begin{figure}[t]
\begin{center}
\includegraphics[scale=0.35, angle=0]{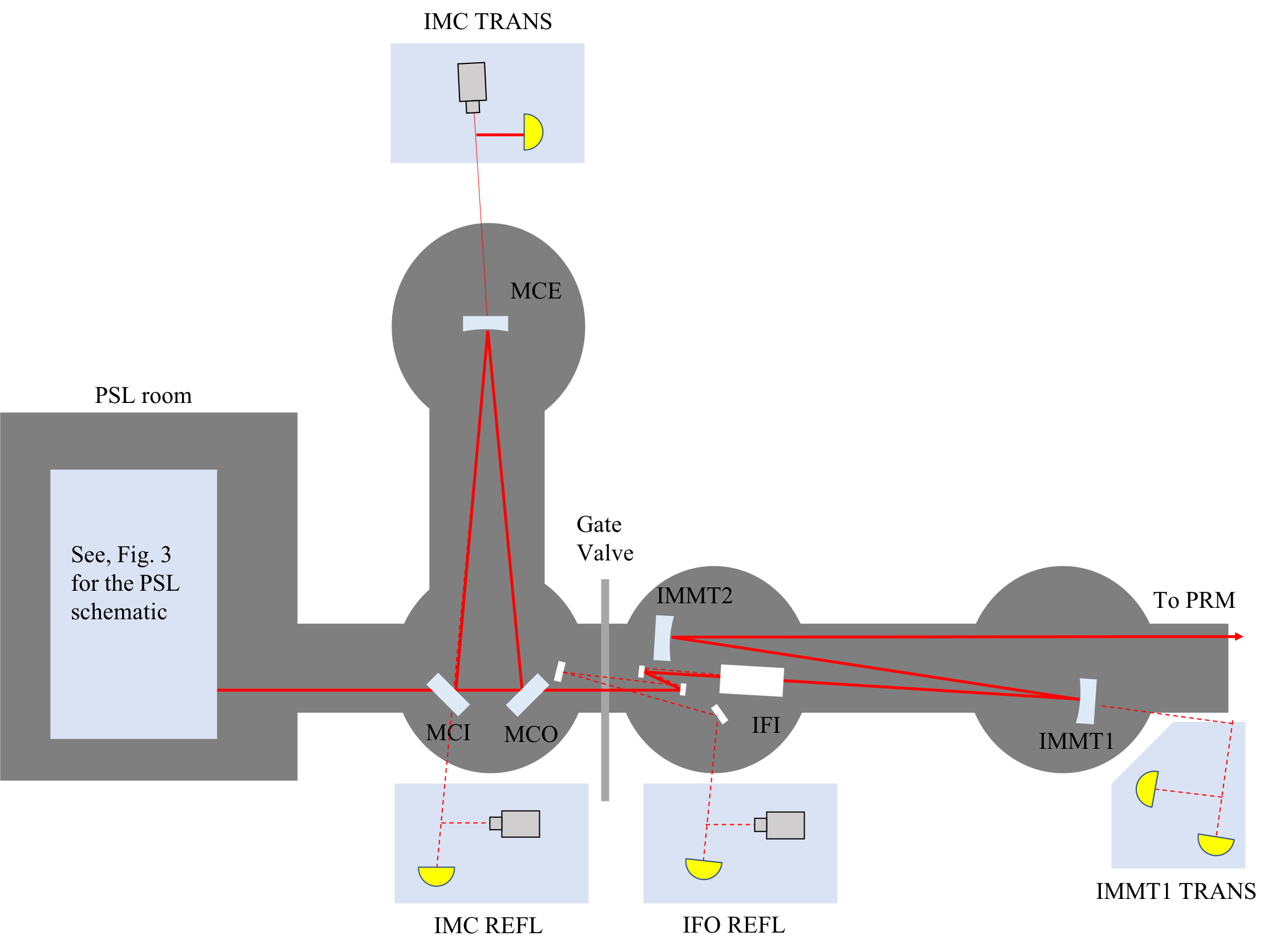}
\caption{\label{config:IOO1}Overview of the input optics for the KAGRA detector.
The laser source is placed in the PSL room.
The laser is transmitted through the IMC
and IFI, and is then reflected
by the first and second IMMTs
before entering the main interferometer.
IMC REFL, IMC TRANS, IFO REFL, and IMMT1 TRANS are the in-air detection benches
to extract signals for the interferometer sensing and controls.
The dashed line is the beam returning from the interferometer.
This beam is separated by the IFI and transmitted to the IFO REFL bench for signal extractions
for the main interferometer.
See, Fig.~\ref{fig:PSL} for the PSL layout.}
\end{center}
\end{figure}

\subsection{Optical configuration of the KAGRA detector during the O3GK observation}
A schematic of the KAGRA laser interferometer is depicted in Fig.~\ref{fig:ifoconfig} \cite{KAGRA:2022fgc, Akutsu:2020his}.
It is based on an L-shaped Michelson interferometer,
with two 3-km Fabry-Perot arms to enhance the GW signals.
During O3GK,
a power-recycling cavity was implemented in addition to the arm cavities
to enhance the effective laser power in the interferometer. This configuration is called
a power-recycled Fabry-Perot Michelson interferometer,
whereas the standard design in the second-generation
ground-based GW detectors, such as aLIGO and AdV,
is the dual-recycled Fabry-Perot Michelson interferometer,
which deploys in additional a signal-recycling cavity~\cite{TheLIGOScientific:2014jea, TheVirgo:2014hva}\footnote{AdV detector was also a power-recycled Fabry-Perot Michelson interferometer in O3.}.


The aim of the input optics system of KAGRA is to provide a low-noise and stable laser field. The requirements to attain the target sensitivity for the input laser beam are: (1) the spatial mode content is the fundamental Gaussian mode (2) the laser spatial mode must match the spatial mode of the power-recycled Fabry-Perot interferometer (3) laser intensity and frequency noise are stabilized and enough low (4) the laser field consists of the carrier and RF sideband fields to obtain interferometer length and angular sensing signals. In addition to the above points, the input optics system serves to (5) prevent the reflected laser field by the power-recycled Fabry-Perot interferometer from returning to the laser source to protect the laser system.

(1) The spatial mode content is cleaned by two mode-cleaning cavities, the 2-m pre-mode cleaner (PMC) placed in the laser room (Sec.~\ref{sec:PMC}) and 55-m input mode cleaner (\ref{sec:IMC}). (2) The spatial mode matching is done by two suspended curved mirrors, input mode matching telescopes, described in Sec.~\ref{sec:IMMT}.
(3) The stabilization for the laser frequency and intensity is depicted in Sec.~\ref{sec:FSS} and Sec.~\ref{sec:ISS}, respectively. (4) The modulation system will be reported in Sec.~\ref{sec:EOM}. For (5), there is an in-vacuum Faraday isolator between the input mode cleaner and the power-recycled Fabry-Perot interferometer. This system was developed together with LIGO and will be reviewed in Sec.~\ref{sec:IFI}. These sections are organized in the order from laser upstream (the laser source) to downstream (the power-recycled Fabry-Perot interferometer), not in the order of the functions of (1) -- (5).

These input optics subsystems and the main interferometer are operated by
a digital real-time control system imported from LIGO \cite{ref:CDS}.
The digital real-time control system works with synchronized high-speed front-end computers,
analog-to-digital converters, and digital-to-analog converters.
Various sensor signals of the input optics system are transmitted to the digital system
for monitoring and controlling purposes.
Because the digital system operates at a sampling frequency of 64 kHz, which is 
then down-converted to 16 kHz, systems requiring fast controls
(such as intensity and frequency stabilization servos, mentioned in the following sections)
are controlled by analog servos.
Even in those cases, important signals are digitally sampled and transmitted to
the data acquisition system for monitoring and characterizing the systems.

\section{Pre-stabilized laser}\label{sec:PSL}

\begin{figure}[t]
\begin{center}
\includegraphics[scale=0.7]{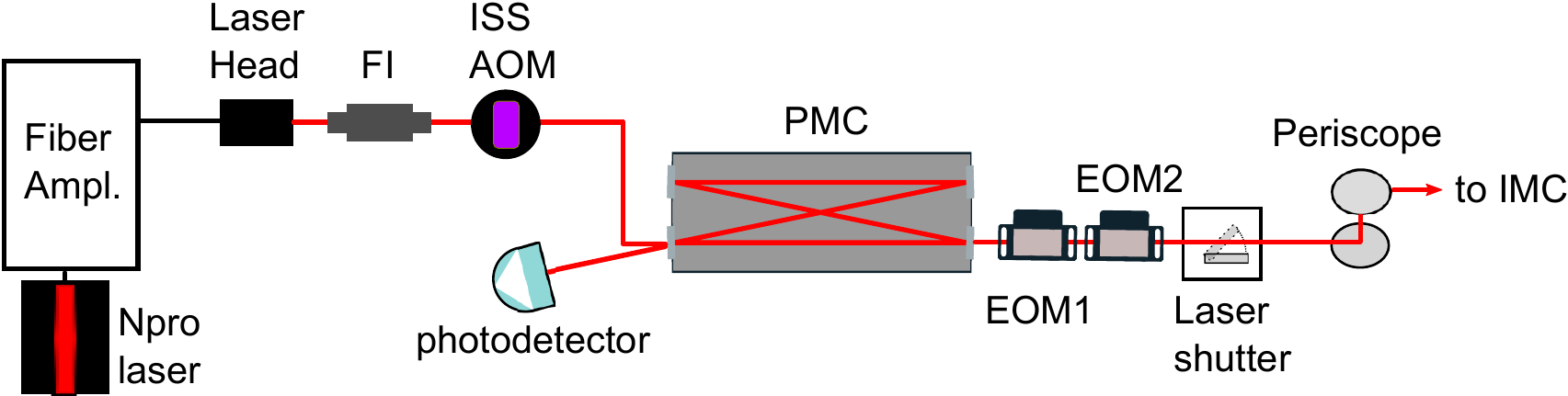}
\end{center}
\vspace{5mm}
\caption{\label{fig:PSL}Simplified PSL layout. The NPRO laser is amplified by the commercial fiber amplifier (shown as Fiber Ampl.) up to 40W. There is a commercial Faraday isolator (FI) just after the laser injection head
to prevent the reflected beam from returning to the laser source. An acousto-optic modulator (shown as ISS AOM) is placed before the pre-mode cleaner (PMC) to work as the actuator for the intensity stabilization loop.
The PMC is a bow-tie-shaped optical cavity used to reject higher-order spatial modes. Two electro-optic modulators (EOMs) are used as phase modulators to extract the IMC and main interferometer signals. 
Each EOM can modulate the laser beam at two RF frequencies with two resonant circuits. EOM1 applies the modulations for the IMC (14 MHz) and f1 (16.88 MHz, the first RF phase modulation for the main interferometer),
and EOM2 applies the modulations for f2 (45 MHz) and f3 (51 MHz). The laser shutter is high-power compatible, originally designed in Albert Einstein Institute for Advanced LIGO.
The periscope raises the beam axis from the height at the PSL table to the height of the input mode cleaner (IMC) in a vacuum chamber.}
\end{figure}

\subsection{The PSL room}

The KAGRA PSL room has a nested structure. The outer structure is a soundproof room of ISO class 4 cleanliness level, maintained by two precision air conditioners with ultra-low particle air filters. The inner structure is a super-clean room of ISO class 1 cleanliness level, maintained by eight open clean benches (KOACH G-1050F, KOKEN Co., Ltd.). During the observation, only one of the two precision air conditioners is operated to maintain silence, low airflow disturbance, and a constant temperature ($\pm$ 0.05 $^{\circ}$C).

\begin{figure}[htbp]
\begin{tabular}{cc}
\begin{minipage}{0.47\hsize}
\begin{center}
    \includegraphics[scale=0.13]{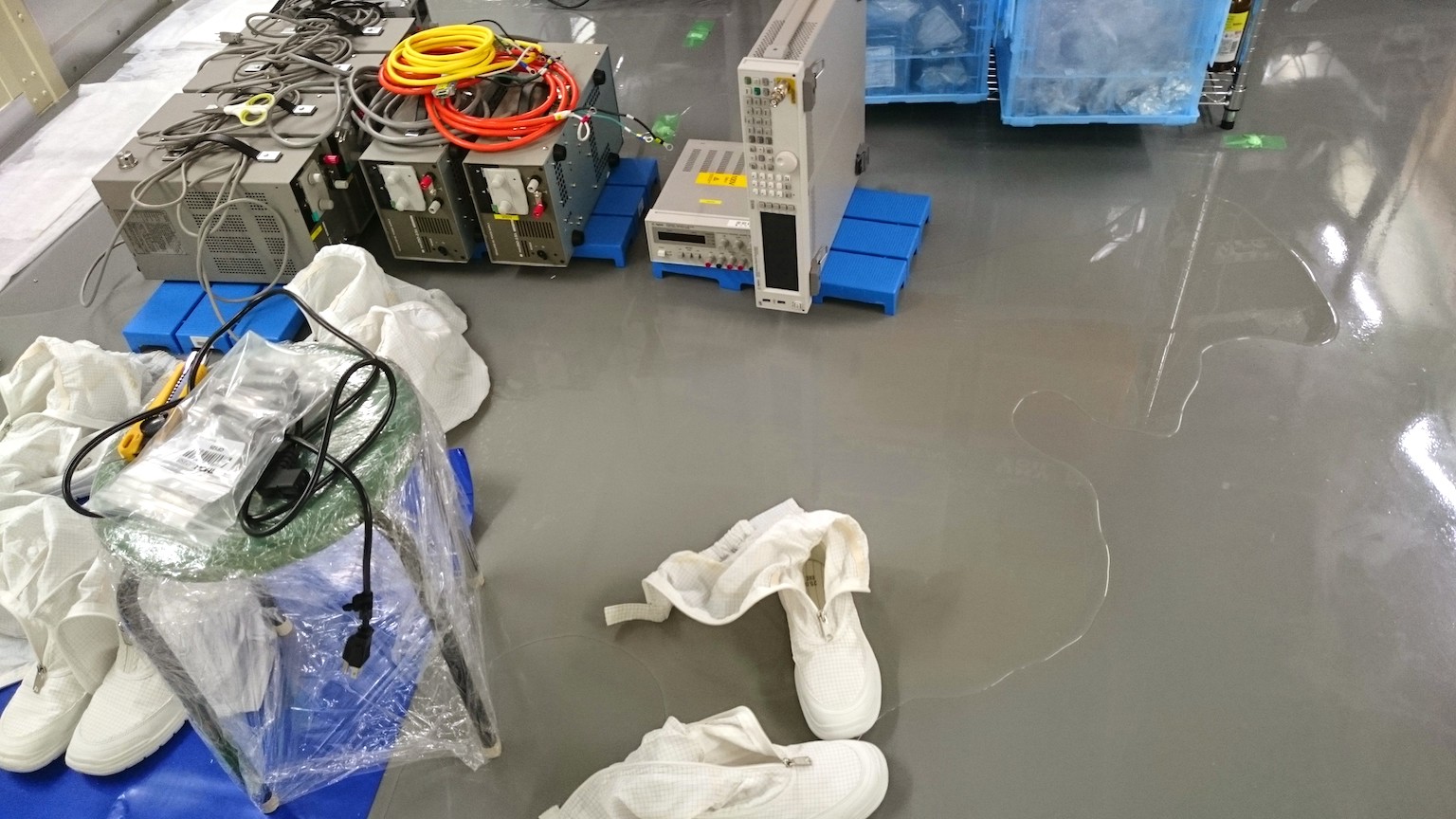}
\end{center}
\end{minipage}
\begin{minipage}{0.47\hsize}
\begin{center}
    \includegraphics[scale=0.13]{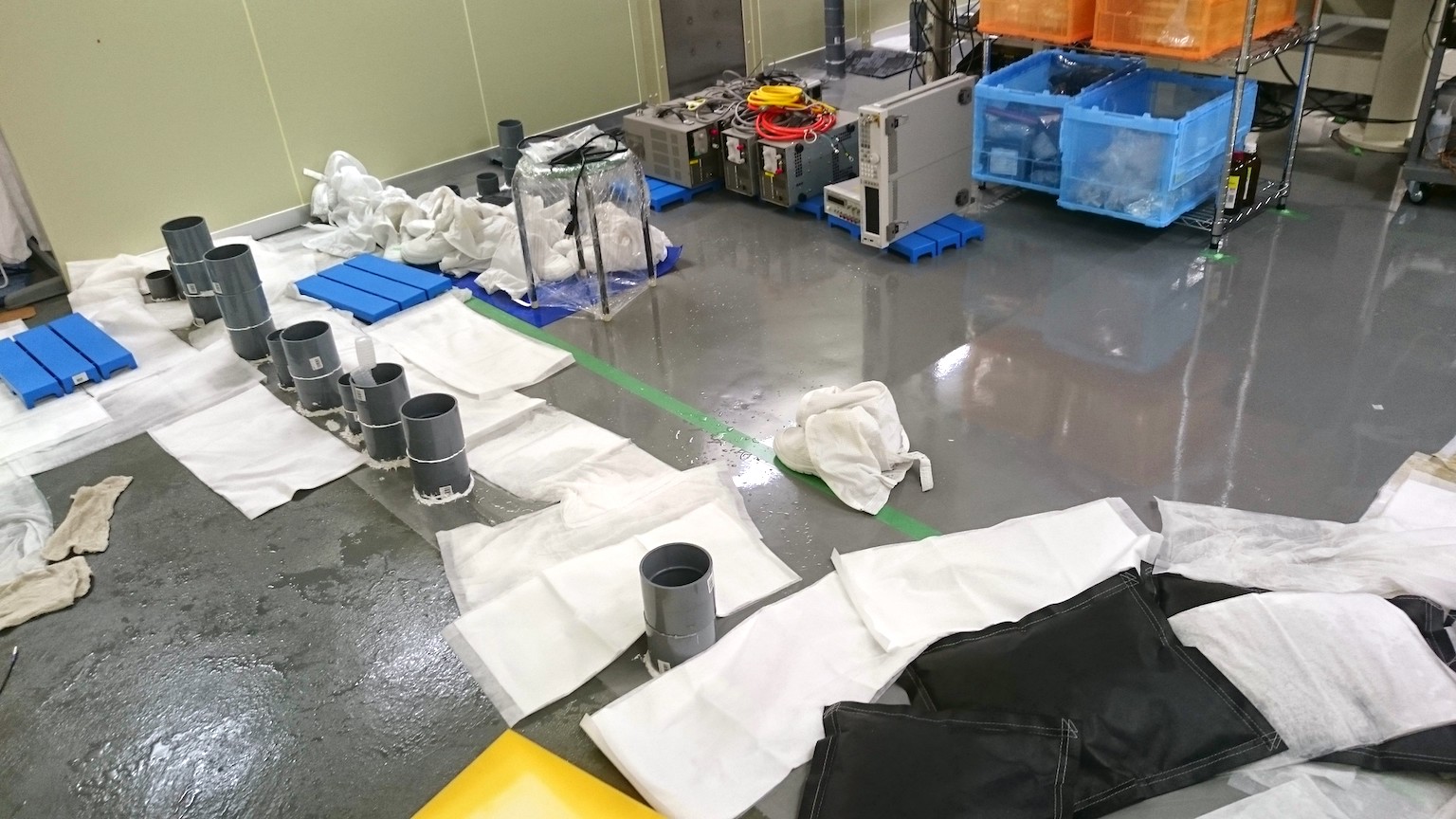}
\end{center}
\end{minipage}
\end{tabular}
\caption{\label{fig:pslwater}Spring water leaked on the floor in the PSL room. (Left) The entire floor of the room was wet. (Right) Pipes were placed at the leaking points on the floor to prevent the water from spreading further. Water absorb sheets are also seen. The legs of the optical table is partially seen on the top right. The photos were taken in March, 2015.}
\end{figure}

Because the PSL enclosure is located in the underground KAGRA tunnel,
in the early phase of the PSL installation, 
the PSL room was affected by underground spring water.
Fig.~\ref{fig:pslwater} shows that the spring water leaked on the floor in the PSL room. 
Underground water leaked from the tunnel floors, walls, and ceilings
through cracks in the surface mortar layers and in the rocks themselves.
Under the PSL enclosure, there are multiple concrete floor frames on the rocks,
separated from each other to insulate ground motions.
Water leaks in the PSL enclosure occurred
through the gaps or cracks in the concrete frames, and
through anchor bolts fixed on both the concrete and rock layers.
To overcome the issue of leaking water, 
drain ditches (20 mm in width) were additionally constructed to lead the water to a sump pit.
The collected water was drained outside of the tunnel by pumps with a flow rate of 0.05 m$^3$/min.
In addition, some cracks were filled using urethane resins.

\subsection{Laser source}
During the observation, KAGRA used a laser system that was
composed of a seed laser and a fiber-laser amplifier with a wavelength of 1064 nm, and a maximum output of 40 W.
We used commercially available lasers: the seed laser was a non-planar ring oscillator (NPRO) (Mephisto 500NEFC, Coherent), and
the fiber-laser amplifier was a PSFA-10 mw-40 W-1064 (Coherent/Nufern).
The amplifier required a water chiller system, which
was placed outside the PSL room to avoid
any noise couplings from the vibration of the chiller.
Between the seed laser and fiber amplifier,
there were an isolator and two fiber-coupled EOMs in series.
The first EOM was used to apply a phase modulation for the signal extraction of the PMC
(see, Sec.~\ref{sec:PMC}); the second EOM was a broadband modulator used as one of the actuators
for the frequency stabilization system (FSS) (see, Sec.~\ref{sec:FSS}).

The frequency noise measured after the laser amplification
was similar to that of the NPRO without any excess noise
added by the fiber amplifier.
On the other hand, the intensity noise of the amplified light was almost 10 times larger than that of the NPRO at approximately 100 Hz. This level is comparable to that of a high-power solid-state laser that has been developed for GW detectors. The laser intensity is stabilized independently from the laser system downstream,
which will be shown in Sec.~\ref{sec:ISS}.

\subsection{Pre-mode cleaner}
\label{sec:PMC}
\begin{figure}[t]
\begin{center}
\includegraphics[scale=0.34]{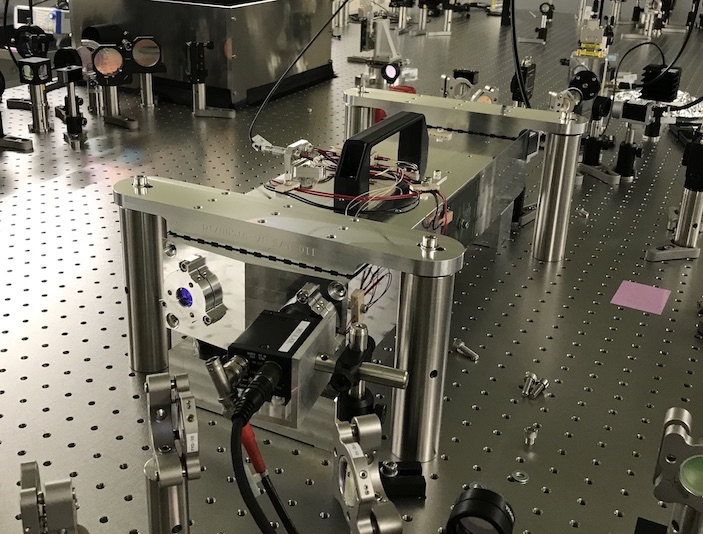}
\caption{\label{fig:PMC}PMC installed on an optical table in the PSL room.
Four mirrors are mechanically attached to a spacer to form
a bow-tie-shaped optical resonator. The length is controlled by a PZT actuator attached to one mirror,
and a thermal heater attached to the sides of the spacer.}
\end{center}
\end{figure}

Spatial modes of the laser field are filtered by the PMC placed
on the optical bench in the PSL room as sketched in Fig.~\ref{fig:PSL}.
The PMC is a bow-tie-shaped cavity with a round-trip length of 2.02 m, shown in Fig.~\ref{fig:PMC}.
Its purpose is not only to filter out higher-order spatial modes
but also to reduce beam jitter at high frequencies.
The KAGRA PMC was designed and built by LIGO \cite{ref:aLIGOPSL, PMCref}.
The cavity length is controlled using the PDH method \cite{Drever:1983qsr};
thus, the laser field's fundamental mode (TEM00 mode) 
resonates in the cavity by a feedback control loop.
An RF phase modulation 
is applied by a fiber-coupled phase modulator (NIR-MPX-LN series, iXblue Photonics)
at the laser source to produce the PDH signal.
The control loop has two types of actuators: a piezoelectric (PZT) actuator attached to one of the four one-inch mirrors
for fast control, and two thermal actuators attached to the spacer
made of 6061 aluminum alloy for slow control.
The free spectral range (FSR) was measured as 147.350 $\pm$ 0.008 MHz,
and the finesse was measured as 121 $\pm$ 2,
whereas the design values for the FSR and finesse are 148 MHz and 124, respectively.
The unity gain frequency of the control is approximately 1 kHz.
The operation was fully automated by the Guardian interferometer automation system \cite{GraefRollins:2016xhy}.
During the O3GK observation, the duty factor of the PMC was 99.5\%.

\subsection{Modulation system}
\label{sec:EOM}
For the sideband generation,
two high-power compatible EOMs installed on the optical bench in the PSL room are used.
Phase modulation is applied at four RF frequencies, $f_{\rm IMC} = 13.78$ MHz, $f_{\rm 1}=16.991$ MHz, $f_{\rm 2} = 45.016$ MHz, and $f_{\rm 3} = 56.270$ MHz. 
The sideband fields at $f_{\rm 1}$, $f_{\rm 2}$, and $f_{\rm 3}$ pass through the IMC to be used for the lock acquisition of the main interferometer. $f_{\rm 3}$ is a non-resonant sideband and is planned for use during a particular phase of the lock acquisition \cite{Aso:2013eba}.
The modulation indices for $f_{\rm 1}$ and $f_{\rm 2}$ sidebands
are measured to be 0.22 and 0.23 rad, respectively.

The EOMs consist of MgO-doped stoichiometric lithium tantalate (SLT)
crystals. MgO-doped SLT crystals are known for their high power
compatibility, small absorption, and small thermal lensing. The EOM
design is based on the aLIGO EOMs, and was produced with assistance from the University of Florida (UF). To avoid etalon interference effects, we choose to wedge the faces of the SLT crystal. The crystal faces are also anti-reflection-coated, with less than 0.1\% remaining reflectivity. The number of modulator crystals is reduced from two to one using two separate pairs of electrodes for two different modulation frequencies to reduce the optical losses. Experiments showed that these EOMs are suitable for KAGRA, owing to low thermal lensing.

\section{Input mode cleaner}
\label{sec:IMC}

\begin{figure}[t]
\begin{center}
\includegraphics[scale=0.18]{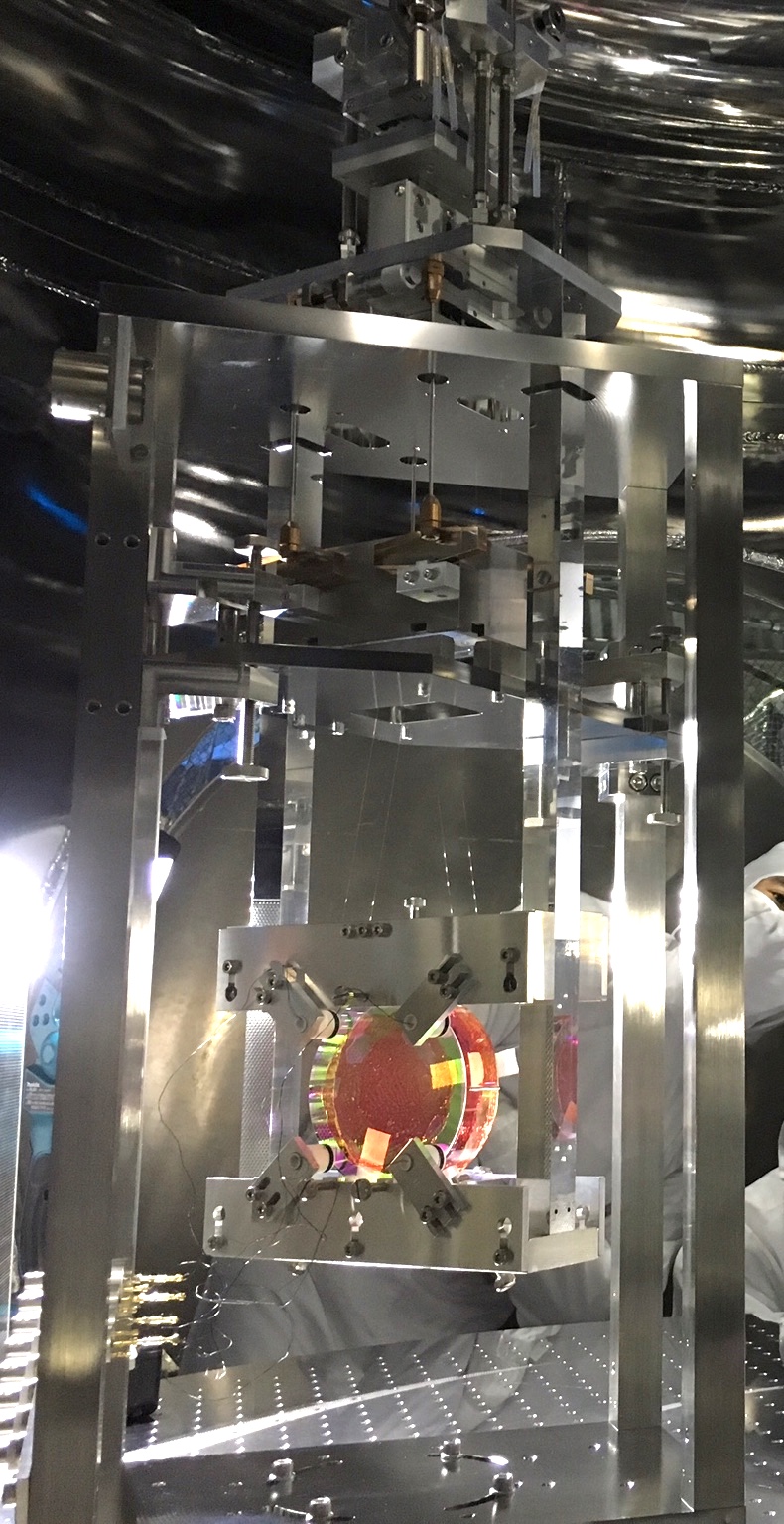}
\caption{\label{fig:typeC}MCO suspension with the mirror, which is a double pendulum
based on the suspension design used for Tama 300. 
 The pink color on the mirror surface
is owing to a polymer coating that temporarily protects the surface.}
\end{center}
\end{figure}

The IMC, the second mode cleaning cavity, is a triangular cavity with a round-trip length of 53.3 m. 
It consists of three mirrors, MCI, MCO, and MCE, as depicted in Fig.~\ref{config:IOO1}.
Double pendulums suspend all three mirrors,
the design of which is based on TAMA suspensions \cite{Takahashi:2002cs} to isolate the mirrors from seismic motions.
One of the suspensions, MCO is shown in Fig.~\ref{fig:typeC}.
The suspensions are placed on vibration isolation stacks \cite{Takahashi:2002wc}
in each vacuum chamber for further isolation of the seismic motions.
With this setup, typical root mean square values of the mirror motions are
0.12 $\mu$rad and 0.15 $\mu$rad for pitch and yaw, respectively.
MCE is in one vacuum chamber, and MCI and MCO are in another vacuum chamber,
connected by a vacuum tube.
The diameter of the IMC mirrors is 100 mm, with a thickness of 30 mm, and a wedge of 2.5$^{\circ}$.
The beam waist is 2.3887 mm at the midpoint between MCI and MCO.
Other mirror parameters are summarized in Table \ref{tab:IMC}.
The duty factor of the IMC was more than 97.4\%
during O3GK.

\begin{table}[b]\centering
\begin{tabular}{llll}
\hline
Mirror & RoC & HR loss & Transmission \\
\hline
MCI & Flat ($>$ 100 km) & $<$ 100 ppm & 6000 ppm ($\pm$ 100 ppm) \\
MCO &Flat ($>$ 100 km) & $<$ 100 ppm & 6000 ppm ($\pm$ 100 ppm)\\
MCE &37.3 m ($\pm$0.1 m) & $<$ 100 ppm & $<$ 200 ppm\\
\hline
\end{tabular}
\caption{\label{tab:IMC}Optical parameters of the IMC mirrors.}
\end{table}

The IMC serves as the reference for the frequency stabilization system.
It will be discussed in Sec.~\ref{sec:FSS}.

\section{Input Faraday isolator}
\label{sec:IFI}

\begin{figure}[t]
\begin{center}
  \includegraphics[width=0.45\textwidth]{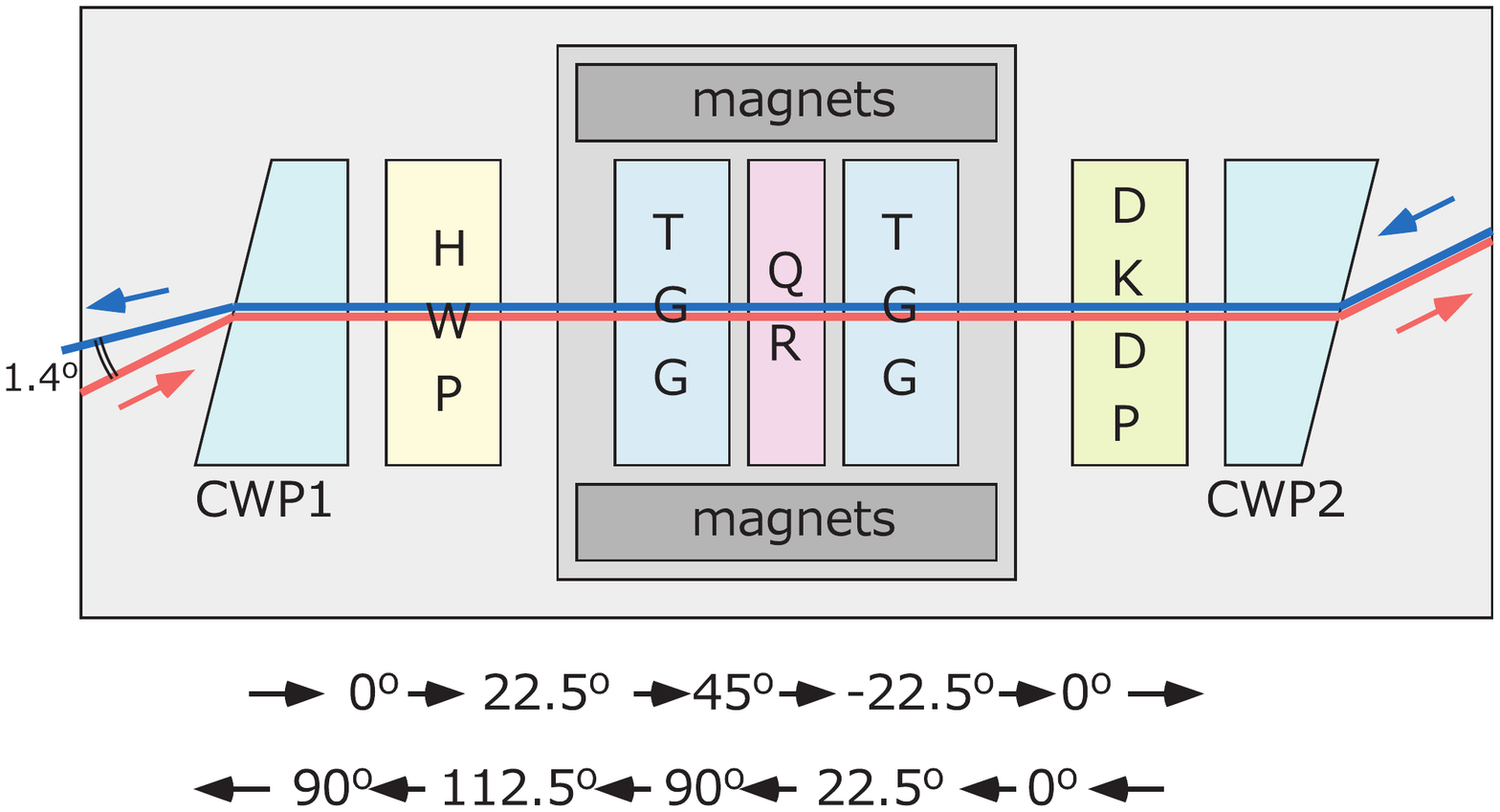}
  \raisebox{7mm}{\includegraphics[width=0.45\textwidth]{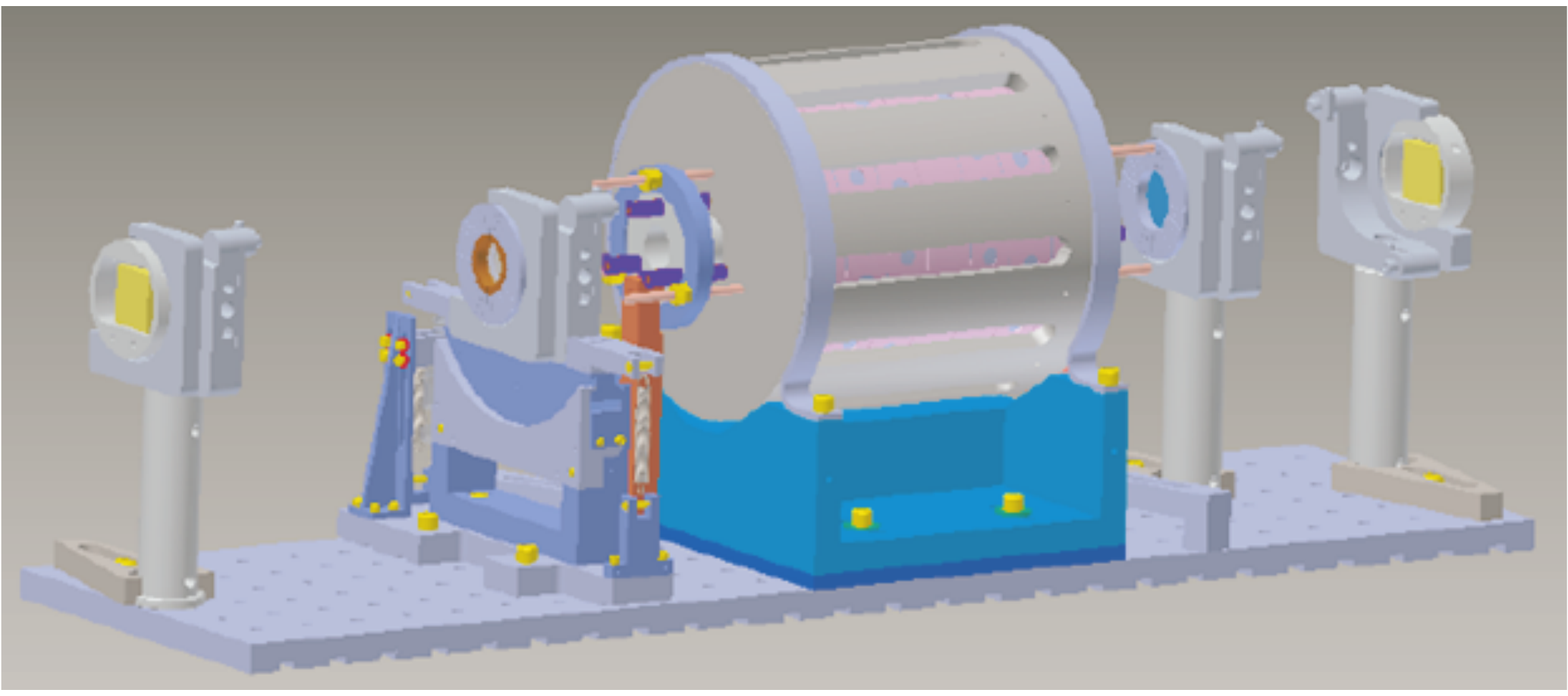}}
\vspace{-3cm}
\caption{\label{fig:overview} {\it Left}: Schematic of the IFI and the polarization angle of the beam behind each optical component. The HWP and QR rotate the polarization of the beam, propagating left to right, by $22.5^\circ$ and $-67.5^\circ$, respectively. For the right-to-left beam propagation, the rotations are $-22.5^\circ$ and $67.5^\circ$, respectively. Each TGG rotates the polarization by $22.5^\circ$, regardless of the propagation direction. {\it Right}: Overview of the IFI.}
\end{center}
\end{figure}

The IFI diverts interferometer-reflected laser light, protecting the laser source. The vacuum-compatible high-power isolator was initially developed at UF for aLIGO~\cite{Mueller:2016hex}; we customized it for KAGRA. The left and right panels of Fig.~\ref{fig:overview} show a schematic and overview of the IFI, respectively. It consists of two calcite-wedge polarizers (CWP1 and CWP2), an HWP, two terbium gallium garnet (TGG) magneto-active crystals, a quartz rotator (QR), and a deuterated potassium dihydrogen phosphate (DKDP) crystal for thermal compensation. Vertically-polarized light, injected from the left side of the figure, is incident on CWP1 at an angle of $7.9^\circ$. The light is transmitted through the other optics, exiting as vertically-polarized light. Reflected light returns to CWP1 with horizontal polarization. The wedge refracts the horizontally-polarized light by an angle that is $1.4^\circ$ smaller, departing from the path of the input beam and transmitting it to a photodetector. The horizontally-polarized reflected beam contains information about mirror locations and alignments; the detected signal is used in the interferometer controls.

According to Ref.~\cite{Mueller:2002du}, thermal lensing in TGG decreases the fundamental mode intensity by 32\,\% for a 150\,W laser, even if the mode-matching is re-optimized using a spherical lens. The DKDP crystal~\cite{Zelenogorsky:2007xrp} in the KAGRA IFI provides a negative thermal lens, which is essential for compensation. Thermal depolarization also occurs in TGG at high powers. As described in Ref.~\cite{Khazanov:2000riw}, two TGG crystals (each of which rotates the polarization by $22.5^\circ$) sandwiching a reciprocal QR provide compensation.

 All the IFI components are ultra-high vacuum compatible. The magnet rings were vacuum baked and then assembled at UF. Each magnet ring consists of 8 wedged sectors. The complete magnet uses 7 magnet rings in a Halbach array. The magnetic field decreases rapidly with increased distance from the IFI; thus, suspensions near the IFI are not influenced. Two TGG crystals and the QR are contained in an aluminum tube that is cantilevered into the magnet. The HWP is mounted on a motorized stage, allowing polarization rotation adjustment. Note that the rotation angle to maximize the transmission, and that to maximize the isolation ratio, can be different. 
The DKDP thickness was 3.0\,mm. The optimal thickness derived from the absorption in the TGG crystals is 3.2\,mm or 3.3\,mm; however, we had to select either 3.0 or 3.5\,mm.

\begin{figure}[t]
\begin{center}
\includegraphics[scale=0.4]{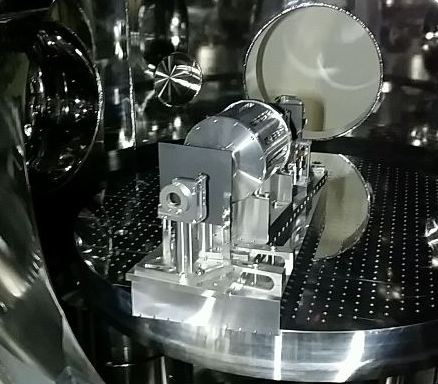}
\caption{\label{fig:IFIphoto}Faraday isolator installed in a vacuum chamber at the KAGRA site.}
\end{center}
\end{figure}

The KAGRA IFI was assembled and aligned in the ISO class 1 laser room in fall 2014. First, the TGG crystal was positioned in the magnet to provide a design polarization rotation angle of ($22.5^\circ$). The measured values were $22.54^\circ \pm 0.01^\circ$. Second, the CWPs were aligned to maximize the extinction ratio. The design value is 50\,dB; we measured a value of 43\,dB. In December 2014, we measured transmission of 97\% and an isolation ratio of 44.7\,dB. 

In December 2015, the IFI was installed on an aluminum spacer in the IFI chamber, as shown in Fig.~\ref{fig:IFIphoto}. The spacer height had an error of 1\,cm, corrected by IMMT1/2. The fine alignment of the beam to the IFI employs two motorized steering mirrors. The unsuspended IFI and mirrors are located on the seismic isolation stack in the IFI chamber, where the seismic noise levels were calculated to be less than the requirement at frequencies above 10\,Hz, including a safety factor of 10. The reflected beam from the main interferometer, which leaves the IFI in horizontal polarization, is shown in Fig.~\ref{config:IOO1}, being transmitted to the control-signal extraction port (IFO REFL). 

\section{Input mode-matching telescopes}\label{sec:IMMT}

The geometrical mode of the input beam injected into the main interferometer must match the mode
of the main interferometer.
%
To convert the beam at the IMC output,
two curved mirrors are placed between the IMC and PRM
to serve as a telescope.
The two mirrors are called IMMT1 and IMMT2
with optical parameters shown in Table. \ref{tab:IMMT}.
They are also used to steer the beam toward the main interferometer.
They are suspended by double pendulum suspensions,
which are of the same type as the IMC suspensions;
the suspension cages are placed on the stacks.
The cages of the IMMT suspensions were covered by
shields made of stainless steel (SUS304) plates, coated by Solblack
(low-reflection and high-absorption black plating; Asahi Precision Co., Ltd.),
to protect the thin suspension wires
from exposure to a high-power beam directly reflected by the PRM
during lock acquisitions of the main interferometer.

\begin{figure}[t]
\begin{center}
\includegraphics[scale=0.75]{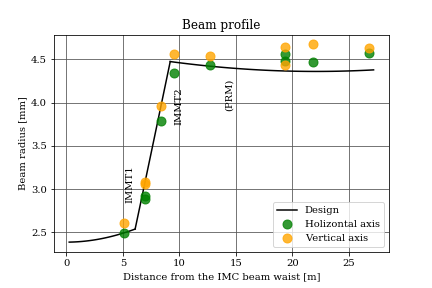}
\caption{\label{fig:IMMT}
Designed beam profile (black) and measured (orange dots for vertical and green dots for horizontal) beam profiles
around IMMT1 and IMMT2. The designed profile is well reproduced.
The beam reflected by IMMT2 was measured without the PRM (position marked in the plot) in the path.}
\end{center}
\end{figure}

The beam size on IMMT1 is designed to be 2.5413 mm,
and to expand to 4.4734 mm at IMMT2; thus, the geometrical mode of the input beam
matches that of the main interferometer.
The measured beam size, shown in Fig.~\ref{fig:IMMT},
is in good agreement with the design values.

The mode matching ratio between the input beam
and each arm was measured as 91\% and 87\% for X and Y arms, respectively, by scanning each arm \cite{Akutsu:2019gxm}.
This ratio is derived by the power in the fundamental mode
and power in the second-order HOM generated by spatial mode mismatches. Note that a fraction of the measured second-order HOM may be generated by inhomogeneities of the input mirrors of the cavities.

A major issue caused by mode mismatch between the input beam
and the arms, i.e., the main interferometer, is a degradation
of signal-to-noise ratio of the length sensing signals at the reflection port
of the interferometer because the mismatched beam is reflected back to the reflection port.
Nominally, signals for common mode motion of the two arm lengths (CARM) are designated to be extracted at this port.
Because the mode-mismatched light does not contribute to
length signal generations, and is rejected to the reflection port, the amount of junk light is increased at the reflection port. On the other hand, with a poor mode-matched beam, less laser power in the main interferometer is circulated, and the power of the length sensing signals is reduced. As a consequence, the mismatched input beam
results in degradation of the signal-to-shot-noise ratio
at the reflection port. This effect, due to the measured mode mismatch, is under investigation.




\begin{table}[t]\centering
\begin{tabular}{lllll}
\hline
Mirror & RoC & HR loss & Transmission & Thickness \\
\hline
IMMT1 & -8.910 m ($\pm$ 0.009 m) & $<$ 1000 ppm & 0.03\% (AOI 2.59 deg) & 52 mm \\
IMMT2 & 14.005 m ($\pm$ 0.238 m) & $<$ 1000 ppm & 0.028\% (AOI 2.59 deg) & 53 mm\\
\hline
\end{tabular}
\caption{\label{tab:IMMT}Optical parameters of the IMMTs. Each mirror has a diameter of 100 mm, and wedge of 2.45$^{\circ}$. AOI: Angle of incidence.}
\end{table}

\section{Laser Noise contributions to the detector sensitivity}\label{sec:noise}

\subsection{Frequency stabilization loop}\label{sec:FSS}

Laser frequency noise is a common noise source for both the X and Y arms; therefore, in an ideal case, it is canceled at the GW channel
at the anti-symmetric port of the interferometer. However, in reality, it couples with the GW channel through practical asymmetries of the interferometer, such as an imbalance of the arm finesse.

\begin{figure}[t]
    \begin{tabular}{cc}
        \begin{minipage}{0.5\hsize}
            \begin{center}
            \includegraphics[scale=0.43]{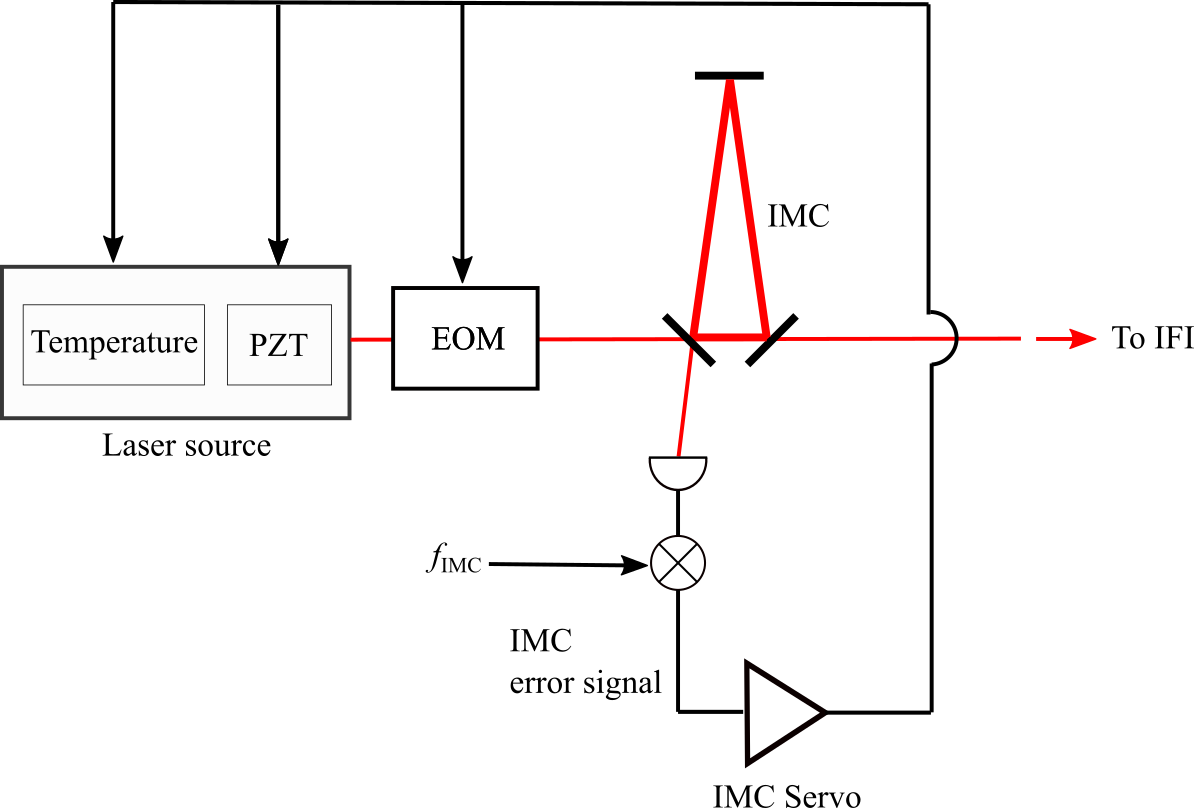}
            \end{center}
        \end{minipage}
                    \hspace{-5mm}
        \begin{minipage}{0.5\hsize}
            \begin{center}
            \includegraphics[scale=0.32, angle=0]{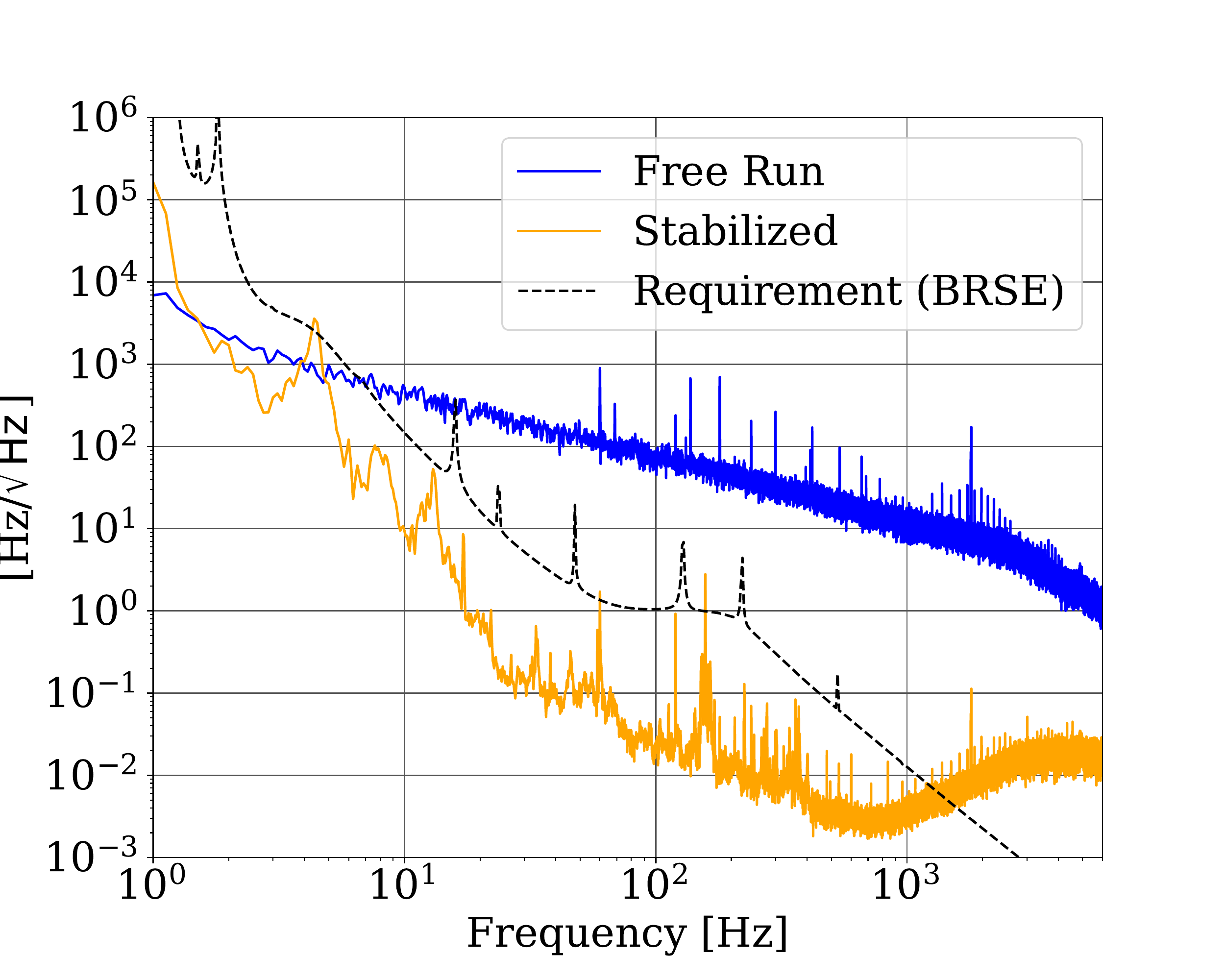}
            \end{center}
        \end{minipage}
    \end{tabular}
    \caption{\label{fig:nestloop}(Left Panel) Control loop schematic of the frequency stabilization with the IMC. The control signals are fed back to the broadband EOM, the laser frequency, and the temperature of the laser crystal. (Right panel) Frequency noise at the IMC output. Compared with free run (blue), the frequency noise is stabilized for about three orders of magnitude at 1 kHz.}
\end{figure}

The final requirement for frequency stability at the IMC output ({\it i.e.},
the input to the main interferometer) to accomplish the design sensitivity \cite{Aso:2013eba} is  discussed in Sec.~4.2.2 in Ref.~\cite{ref:MIFdoc}.
It depends on the interferometer configuration. For example, for a broadband resonant-sideband extraction (BRSE) configuration,
the required frequency noise level is shown as a dashed black line in the right panel of Fig.~\ref{fig:nestloop} 
under the assumptions of an arm finesse asymmetry of 1.5\% between the X and Y arms and a safety factor of 10 to the final sensitivity. Because the target strain sensitivity was significantly lower than the final sensitivity (approximately two orders of magnitude higher at 500 Hz) in O3GK, the requirement for the frequency noise level was also relaxed.

The laser frequency is stabilized using the IMC length as a frequency reference.
The phase modulation at 13.78 MHz is applied at the PSL table,
and the error signal of the IMC length is obtained from the reflection port of the IMC,
using the PDH method \cite{Drever:1983qsr}.
As shown in the left panel of Fig.~\ref{fig:nestloop}, the control signals are fed back
to the laser frequency, actuated by a broadband EOM (above 15 kHz)
and laser PZT actuator (between 15 kHz and 0.1 Hz), through an analog servo board.
In the low-frequency range, below 0.1 Hz, the control signals are fed
back to the laser crystal temperature.
With these control loops, a control bandwidth of 130 kHz was achieved.
The stabilized frequency noise level at the IMC output, as measured from the CARM error signal, is plotted in the right panel of Fig.~\ref{fig:nestloop}. Although the frequency noise level did not meet the requirement above 2 kHz, it was sufficient for O3GK. Below 2 kHz, the frequency noise level was sufficiently below the final requirement in the BRSE.




When the power-recycled Fabry-Perot interferometer is operated, the control loop for the frequency stabilization is nested
with CARM ($L_{\rm CARM}=\frac{L_x+L_y}{2}$,
where $L_x$ and $L_y$ are the lengths of the X and Y arms, respectively),
because it is a better frequency reference at the observation frequency band.

The estimated impact of frequency noise on the detector sensitivity is discussed in Sec.~\ref{sec:noise}.

\subsection{Intensity stabilization system}\label{sec:ISS}

Because fluctuations of the laser intensity contaminate the GW signals through practical asymmetries between the two arms, the laser intensity must be stabilized. For O3GK, our target for relative intensity noise (RIN) was $1\times10^{-7} /\sqrt{{\rm Hz}}$ at several tens of Hz.

The intensity stabilization system (ISS) for the laser was developed to meet this requirement. The system is based on negative feedback using an AOM as an actuator of the laser intensity, where excess power is removed as the first-order diffracted beam from the AOM. The AOM was placed upstream of PMC in the PSL room, as shown in Fig.~\ref{fig:PSL}.

There is a photodetector at the transmission port of the IMMT mirror 1, shown as IMMT1 TRANS in Fig.~\ref{config:IOO1}.
This pickoff port is approximately 13 m downstream of the output of the PSL room.
The detection system consists of a focusing lens, half-wave plate (HWP), beam splitter, and two photodetectors. The vertical polarization of the laser beam is transformed into a horizontal polarization using the HWP, and is split into two beams of nearly equal powers by the beam splitter; it is then focused near the surface of the InGaAs PIN photodiodes (Excelitas technologies C30642). 

The output signal from one of the two PDs is transmitted to a signal conditioning filter circuit in the PD head, and it is then compared with a low-pass filtered reference direct current voltage. The differential signal between them is transferred to the servo-electric circuits and fed back to the AOM to stabilize the intensity noise. 
%

\begin{figure}[t!]
\begin{center}
\includegraphics[scale=0.4]{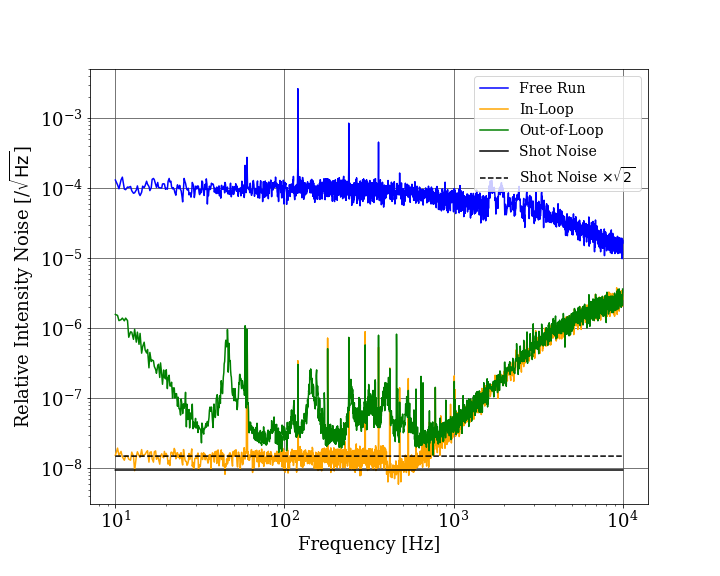}
\end{center}
    \caption{\label{fig:ISS}Relative intensity noise of the laser. The laser power was 5.6\,W.}
\end{figure}

Without the intensity stabilization, the RIN level is as high as $10^{-4}/\sqrt{{\rm Hz}}$ and is degraded through the IMC by about an order of magnitude.
The cause of this degradation is not understood.
By closing the loop, the system successfully reduces the RIN by approximately three orders of magnitude to below the target value of $10^{-7} /\sqrt{{\rm Hz}}$ in the frequency range between 30\,Hz and 1\,kHz, as shown in Fig.~\ref{fig:ISS}. The RIN of the shot noise with an incident power of 4.8 and 5.6\,mW for in- and out-of-loop PDs, respectively, is approximately $1.4\times10^{-8} /\sqrt{{\rm Hz}}$. There are several spikes of 60\,Hz, owing to the use of a commercial power source and its harmonic frequencies. The noise at lower than 30\,Hz, and a broad spectrum around 46\,Hz originates from the beam jitter. Some coherence was observed between the angular fluctuations in the PSL room and the RIN. The structure at higher than 1\,kHz is owing to the shortage of a gain in the servo loop.

\subsection{Projections to the Sensitivity}
\begin{figure}[t]
\begin{center}
\includegraphics[scale=0.45]{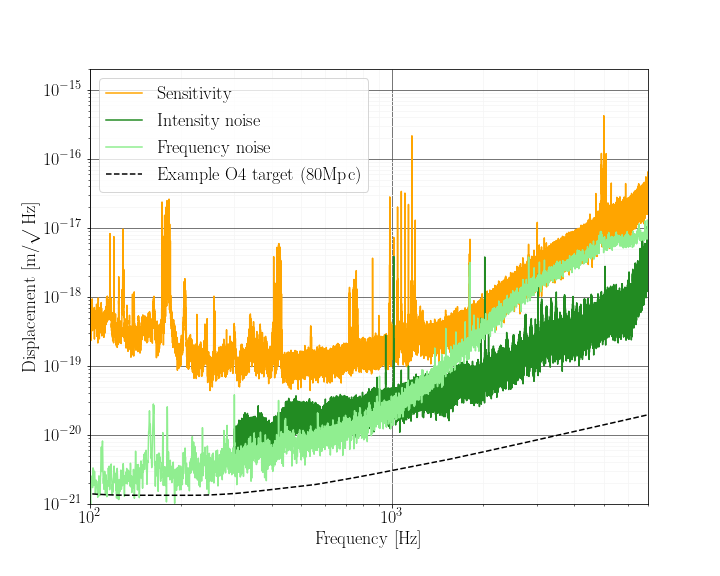}
\caption{\label{fig:noise} Typical displacement sensitivity curve during O3GK (orange) with intensity noise (dark green) and frequency noise (light green, in-loop) projections above 100 Hz. For the O3GK observation, the intensity noise did not contaminate the detector sensitivity, whereas the frequency noise may limit the sensitivity above a few kHz. To achieve the target sensitivity of the next observing run (black dashed curve),
further stabilization both for intensity and frequency will be necessary.}
\end{center}
\end{figure}

Fig.~\ref{fig:noise} shows the estimation of the intensity
and frequency noise contributions in the GW sensitivity spectrum (also presented in Ref. \cite{KAGRA:2022fgc}).
While the intensity noise (dark green curve) did not limit
the GW sensitivity (orange curve),
the frequency noise (light green curve) limited the sensitivity above a few kHz.
The estimation of the intensity noise contribution was derived by measuring the RIN at the ISS photodetector,
and multiplying it by a measured transfer function from the feedback point of the ISS loop
to the detector sensitivity signal, calibrated in m/$\surd$Hz.
It is not plotted below a few hundred Hz because the coherence
of the transfer function was low.
The frequency noise contribution was estimated in a similar manner.
In this case, with the main interferometer, the FSS loop was nested with the CARM loop.
This residual frequency noise projection is in-loop
with a unity gain frequency of approximately 20 kHz.
Although the noise budget of the FSS loop has not yet been closely investigated,
we assume that the electrical noise of the servo circuits in the loop was
not suppressed sufficiently and limited the residual noise of the FSS loop.
Note that the transfer functions of both the intensity and frequency noise estimations
exhibited very low coherence between the injected and measured signals below 300 Hz,
and the noise estimation may not be accurate in the lower frequency region.


\subsection{Future improvements}\label{sec:plans}
The intensity and frequency noise stabilization must be improved for future observing runs in the broad frequency region.

For the ISS, the final target for RIN is $2 \times 10^{-9}/\sqrt{\mathrm{Hz}}$ at several tens of Hz.
To achieve this, the monitoring photodetectors are planned to be moved into a vacuum chamber to reduce environmental noises such as air disturbances. Also, the number of monitoring photodiodes will be increased from two to four because a higher laser power received at the monitoring photodetectors can reduce the shot noise. Furthermore, a beam jitter control loop will be implemented to prevent beam position fluctuation on the monitoring photodetectors.

For the FSS, further loop optimization and noise hunting will be necessary. For instance, below 1 kHz, the control loop is limited by the dark noise of the CARM PD. more power at the PD will improve the signal-to-noise ratio because the CARM signal is proportional to the light power at the PD.
Also, during O3GK, the CARM PD was in the air. It will be relocated to a vacuum chamber to avoid air turbulence.
Moreover, we observed that the frequency noise contribution to the sensitivity fluctuated and significantly changed by approximately one order of magnitude, strongly depending on the alignment status of the main interferometer. An active global angular control scheme of the main interferometer will be useful in reducing this effect.

An asymmetry between the X and Y arms, and the contrast defect in the Michelson interferometer, are the origins of intensity and frequency noise couplings to the gravitational-wave channel. An asymmetry in the cavity finesse was found to be an order of magnitude larger than the required value, as reported in Ref. \cite{KAGRA:2022fgc}. Also, birefringence and inhomogeneity in input test mass (IX and IY) substrates created larger contrast defects, resulting in larger laser noise couplings \cite{Michmura:2019}. Further investigations are underway to reduce these noise couplings.

Furthermore, beam dumps and baffles to eliminate ghost beams and to prevent scattered lights, respectively, are planned to be installed around the IFI, IMMT1, and IMMT2. The scattered lights are one of the anticipated noise sources in the next observations at better sensitivity.


\section{Conclusion}

We have described the design and performance of the input optics systems used in O3GK.
The EOMs, IMC, IFI, and IMMTs were successfully installed and operated during the observation. The laser intensity and frequency stabilization system attained the required noise level for the observation.

\section*{Appendix}\label{sec:abb}
Table \ref{tab:abbr} lists the acronyms used in this paper.

\begin{table}[htbp]\centering
\begin{tabular}{ll}
\hline
AdV & Advanced Virgo\\
aLIGO & Advanced LIGO\\
AOM & acousto-optic modulator \\
BNS & binary neutron star inspirals\\
BRSE & broadband resonant-sideband extraction \\
BS & beam splitter\\
CARM & common mode arm length \\
CWP & Calcite-wedge polarizers \\
DKDP & deuterated potassium dihydrogen phosphate \\
EOM & electro-optic modulator \\
EX & end test mass X\\
EY & end test mass Y\\
FSS & frequency stabilization system\\
HOM & higher order (spatial) mode\\
HWP & half wave plate \\
IFI & input Faraday isolator \\
IMC & input mode cleaner \\
IMMT & input mode-matching telescope\\
IO & input optics \\
ISS & intensity stabilization system\\
IX & input test mass X\\
IY & input test mass Y\\
OFI & output Faraday isolator\\
OMC & output mode cleaner\\
OMMT & output mode-matching telescope \\
PD & photodetector \\
PMC & pre-mode cleaner \\
PRM & power-recycling mirror \\
PSL & pre-stabilized laser \\
QR & quartz rotator \\
RIN & relative intensity noise \\
SLT & stoichiometric lithium tantalate\\
SRM & signal-recycling mirror \\
TGG & terbium gallium garnet \\
\hline
\end{tabular}
\caption{\label{tab:abbr}Abbreviations used in this article.}
\end{table}


\section*{Acknowledgment}
This work was supported by MEXT, JSPS Leading-edge Research Infrastructure Program, JSPS Grant-in-Aid for Specially Promoted Research 26000005, JSPS Grant-in-Aid for Scientific Research on Innovative Areas 2905: JP17H06358, JP17H06361 and JP17H06364, JSPS Core-to-Core Program A. Advanced Research Networks, JSPS Grant-in-Aid for Scientific Research (S) 17H06133 and 20H05639 , JSPS Grant-in-Aid for Transformative Research Areas (A) 20A203: JP20H05854, the joint research program of the Institute for Cosmic Ray Research, University of Tokyo, National Research Foundation (NRF), Computing Infrastructure Project of KISTI-GSDC, Korea Astronomy and Space Science Institute (KASI), and Ministry of Science and ICT (MSIT) in Korea, Academia Sinica (AS), AS Grid Center (ASGC) and the Ministry of Science and Technology (MoST) in Taiwan under grants including AS-CDA-105-M06, Advanced Technology Center (ATC) of NAOJ, and Mechanical Engineering Center of KEK.

\bibliographystyle{ieeetr}
\bibliography{ref}

\end{document}